\documentclass[a4paper,fleqn]{cas-dc}
\newcommand{\tableme}[1]{Table:\ref{#1}}
\newcommand{\figref}[1]{Fig:\ref{#1}}
\usepackage{hyperref}
\usepackage{tikz}

\usepackage[linesnumbered,ruled,vlined]{algorithm2e}
\usepackage{dirtytalk}
\usepackage{psfrag}
\usepackage{balance}
\newcommand{\redDashedLine}{\textcolor{red}{\rule[0.5ex]{.2em}{0.8pt}\hspace{0.2em}\rule[0.5ex]{.2em}{0.8pt}\hspace{0.2em}\rule[0.5ex]{.2em}{0.8pt}\hspace{0.2em}\rule[0.5ex]{.2em}{0.8pt}}}

\newcommand{\blackDashedLine}{\textcolor{black}{\rule[0.5ex]{.2em}{0.8pt}\hspace{0.2em}\rule[0.5ex]{.2em}{0.8pt}\hspace{0.2em}\rule[0.5ex]{.2em}{0.8pt}\hspace{0.2em}\rule[0.5ex]{.2em}{0.8pt}}}

\usepackage{xcolor}
\definecolor{c1}{HTML}{fc8d59}
\definecolor{c2}{HTML}{ef6548}
\definecolor{c3}{HTML}{d7301f}
\definecolor{c4}{HTML}{b30000}
\definecolor{c5}{HTML}{7f0000}

\SetKwInput{KwInput}{Input}                
\SetKwInput{KwOutput}{Output}              
\usepackage[nocomma]{optidef}
\usepackage{multicol}
\usepackage{subcaption}



\usepackage{multicol} 

\usepackage{nomencl} 

\usepackage{amssymb}
\usepackage{nomencl}
\makenomenclature
\usepackage{etoolbox}
\renewcommand\nomgroup[1]{%
  \item[\itshape
  \ifstrequal{#1}{A}{Sets}{%
  \ifstrequal{#1}{B}{Indices}{%
  \ifstrequal{#1}{C}{Parameters}{
  \ifstrequal{#1}{D}{Variables}
  }}}]}

\usepackage[numbers]{natbib}
\usepackage[printonlyused]{acronym}

\def\tsc#1{\csdef{#1}{\textsc{\lowercase{#1}}\xspace}}
\tsc{WGM}
\tsc{QE}
\tsc{EP}
\tsc{PMS}
\tsc{BEC}
\tsc{DE}
\usepackage{lipsum}

\begin{document}

\nomenclature[A]{$\mathcal{T}$}{Set of all time-steps considered in the optimisation problem.}
\nomenclature[A]{$\mathcal{N}$ }{Set of all \ac{EV} transactions in the chosen sample.}
\nomenclature[A]{$\mathcal{T}_f$}{Set of all time-steps considered in the flexibility request window.}


\nomenclature[C]{$\overline{p}_{n}$ }{The maximum charging/ discharging power of $n^\text{th}$ \ac{EV} $\forall \: n\in \mathcal{N}$.}
\nomenclature[C]{$\underline{p}_{n}$}{The minimum charging/ discharging power of $n^\text{th}$ \ac{EV} $\forall \: n\in \mathcal{N}$.}
\nomenclature[C]{$\overline{e}_{n}$}{The maximum \ac{SOE} of $n^\text{th}$ \ac{EV}$\quad\forall \: n\in \mathcal{N}$.                }
\nomenclature[C]{$\Delta t$}{Duration of one time-step.   }
\nomenclature[C]{$P_t^{base}$}{The aggregate charging power demand of all the \acp{EV} assuming dumb charging at time-step $t$.}

\nomenclature[C]{$\epsilon$}{A small number used as a multiplier for balancing the numerical sensitivity of the objective function.}
\nomenclature[C]{$t_n^a$}{Arrival time of $n^{th}$ \ac{EV}}
\nomenclature[C]{$t_n^d$}{Departure time of $n^{th}$ \ac{EV}}

\nomenclature[D]{$p_{n,t}$}{The charging/ discharging power of $n^\text{th}$ \ac{EV} at time-step $t$; $\forall \: t\in \mathcal{T}\text{ and } n\in \mathcal{N}$.}
\nomenclature[D]{$e_{n,t}$}{The energy charged by $n^\text{th}$ \ac{EV} at time-step $t$; $\forall \: t\in \mathcal{T}\text{ and } n\in \mathcal{N}$.}
\nomenclature[D]{$c^l$}{The capacity limitation power}
\nomenclature[D]{$c^r$}{The redispatch power w.r.t the base profile (unoptimised)}
\let\WriteBookmarks\relax
\def\floatpagepagefraction{1}
\def\textpagefraction{.001}

\shorttitle{Quantifying the Aggregate Flexibility of Electric Vehicles Charging Stations for Dependable Congestion Management Products}

\shortauthors{NK Panda et~al.}

\title [mode = title]{Quantifying the Aggregate Flexibility of Electric Vehicles Charging Stations for Dependable Congestion Management Products - A Dutch Case Study}                   
\tnotemark[1]

\tnotetext[1]{\label{acknowledgement} The research was supported by the ROBUST project, which received funding from the MOOI subsidy programme under grant agreement MOOI32014 by the
Netherlands Ministry of Economic Affairs and Climate Policy and the Ministry of the Interior and Kingdom Relations, executed by the Netherlands Enterprise Agency.}

\author[1]{Nanda Kishor Panda}[orcid=0000-0002-9647-4424]

\cormark[2]

\cortext[2]{Corresponding author}


\ead{n.k.panda@tudelft.nl}


\credit{Conceptualization of this study, Data curation, Methodology, Software, Writing - Original draft preparation}

\author[1]{Simon H. Tindemans}[orcid=0000-0001-8369-7568 ]
\ead{s.h.tindemans@tudelft.nl}
\credit{Conceptualization of this study, Methodology, Writing - editing, Project Administration, Funding Acquisition}

\affiliation[1]{organization={Department of Electrical Sustainable Energy, Delft University of Technology},
    addressline={Mekelweg 4}, 
    city={Delft},
    postcode={2628 CD}, 
    country={The Netherlands}}

\begin{abstract}
Electric vehicles (EVs) play a crucial role in the transition towards sustainable modes of transportation and thus are critical to the energy transition. As their number grows, managing the aggregate power of EV charging is crucial to maintain grid stability and mitigate congestion. This study analyses more than 500 thousand real charging transactions in the Netherlands to explore the challenge and opportunity for the energy system presented by increased charging needs and smart charging flexibility. Specifically, it quantifies the collective ability to provide dependable congestion management services according to the specifications of those services in the Netherlands. In this study, a data-driven model of charging behaviour is created to explore the implications of delivering dependable congestion management services at various aggregation levels and types of service. The probabilistic ability to offer different flexibility products, namely, redispatch and capacity limitation, for congestion management, is assessed for different categories of charging stations (CS) and dispatch strategies. These probabilities can help EV aggregators, such as charging point operators, make informed decisions about offering congestion mitigation products per relevant regulations and distribution system operators to assess their potential. Further, it is shown how machine learning models can be incorporated to predict the day-ahead consumption, followed by operationally predicting redispatch flexibility. The findings demonstrate that the timing of EV arrivals, departures, and connections plays a crucial role in determining the feasibility of product offerings, and dependable services can generally be delivered using a sufficiently large number of CSs. 

\end{abstract}

\begin{keywords}
Electric Vehicle  \sep Congestion \sep Flexibility \sep Market Product \sep Smart charging
\end{keywords}
\maketitle
   \acrodef{ACM}{Authority for Consumers \& Markets}
    \acrodef{BAU}{Business as-usual }
    \acrodef{CP}{Charging Point}
    \acrodef{CS}{Charging Station}
    \acrodef{CSP}{Congestion Service Provider}
    \acrodef{CPO}{Charging Point Operator}
    \acrodef{DSO}{Distribution System Operator}
    \acrodef{EV}{Electric Vehicle}
    \acrodef{EVSE}{Electric Vehicle Supply Unit}

    \acrodef{OCPI}{Open Charge Point Interface}
    \acrodef{SOC}{State of Charge}
    \acrodef{SOE}{State of Energy}
    \acrodef{TOU}{Time of Use} 
\acrodef{LV}{Low Voltage}
\acrodef{RES}{Renewable Energy Source}
\acrodef{SVR}{Support Vector Regressor}
\acrodef{CDR}{Charge Detail Record}
\acrodef{EAN}{European Article Numbering}
\acrodef{HV}{High Voltage}
\acrodef{FCR}{Frequency Containment Reserve}
\acrodef{ICE}{Internal Combustion Engine}
\acrodef{BRP}{Balance Responsible Party}
\section{Introduction} \label{sec:1}
\acp{EV} offer a promising and sustainable alternative to traditional \ac{ICE} vehicles. This has prompted many countries to encourage the adoption of \acp{EV} through subsidies and developing charging infrastructure networks to support this transition. For example, the Netherlands has witnessed significant developments in adopting \acp{EV} and the associated expansion of charging infrastructure. The government aims to reduce carbon emissions by over 95\% by 2050, with a significant portion of these emissions attributed to vehicular traffic. In light of this goal, policymakers have planned to ban the sale of \ac{ICE} powered passenger vehicles by 2030 and plan to decarbonise the entire mobility sector by 2050 through the use of green electricity, green hydrogen, Power-to-X solutions, and bio-fuels~\cite{economische}. In July 2022, \acp{EV} amounted to 31.75\% of the total sales of passenger cars in the Netherlands. Currently (February 2025), over 572,000\footnote{\url{https://duurzamemobiliteit.databank.nl/mosaic/nl-nl/elektrisch-vervoer/personenauto-s}} passengers \acp{EV} are registered in the Netherlands. Of these, more than 13.1\% belong to shared \ac{EV} fleets, which can play a pivotal role in the provision of congestion management services, as will be shown later~\cite{NetherlandsEnterpriseAgency2022ElectricNetherlands}.\par

\subsubsection*{Smart charging flexibility}
With the large-scale adoption of electric vehicles, power
Networks are facing challenges in safely accommodating their
aggregate charging needs. This has resulted in congestion, where
parts of the network risk being overloaded during times of high
demand. Without timely measures, the existing electricity grid in many locations cannot cope with heavy charging demand during peak hours, causing voltage fluctuations and overloading transformers and cables\textcolor{blue}{~\cite{mahmud2023global}}. This has prompted network operators and policymakers to propose near-term solutions that can defer grid reinforcement without compromising the reliability of power networks.\par

In this regard, smart charging (including vehicle-to-grid (V2G)) offers significant potential to mitigate (or exacerbate) network congestion by leveraging the flexibility in charging. Smart charging is the controlled and optimised charging of \acp{EV} with the aim of reducing costs, providing ancillary services to the grid, and increasing the utilisation of existing infrastructure, among other objectives. \ par

Given that \acp{EV} are idle for 95\% of the day, their inherent storage flexibility makes them an attractive option for grid balancing or other ancillary services~\cite{IRENA2019InnovationVehicles}. The services that \acp{EV} offer can range from simple demand response to more complex frequency regulation services\textcolor{blue}{~\cite{sevdari2022ancillary}}. Individual \acp{EV} can participate in demand side management through the retail electricity market by offering services like demand response for financial incentives, optimised charging based on renewable energy generation or providing storage flexibility through bi-directional charging. However, due to uncertain driving patterns of individual users and the large number of \acp{EV} being deployed, the aggregate flexibility of \acp{EV} far surpasses individual potential, emphasising the crucial role of aggregators in effectively utilising \ac{EV} flexibility on a larger scale. An aggregator, in the case of  \acp{EV}, acts as a middleman between the \ac{DSO} and the charging demands of a group of \acp{EV}. For the remainder of this paper, we assume a single \ac{CPO} who also serves as the aggregator of flexibility for the \acp{EV}.

\subsubsection*{Market-based congestion management}
The interaction between the aggregator and the DSO is often facilitated through a market mechanism that matches flexibility demand with flexibility offerings, ensuring the correct balance of value and cost. Depending on the type of ancillary services, there are various ways the market arrangements can be structured~\cite{rancilio2022ancillary}. Market-based solutions can drive aggregate flexibility, making them a viable option for addressing congestion issues~\cite{Huang2015DistributionNetworks}. In this article, we focus specifically on market-based congestion services for aggregate \acp{EV}.  These services offer a potential solution to delay grid reinforcement by addressing the growing demand for \ac{EV} charging.

To fully understand the potential of market-based congestion services, it is crucial to quantify their impact and the associated risks. Doing so can provide valuable insights to network operators, \acp{CPO}, and policymakers. In this article, we show how we can quantify the potential flexibility of \ac{EV} \acp{CS} to provide dependable congestion management products.  Accurate quantification allows these stakeholders to assess the effectiveness of these services in mitigating grid congestion, delaying costly infrastructure upgrades, and integrating more  \acp{EV} into the network. This information can inform strategic decisions regarding grid management, investment in charging infrastructure, and the development of supportive regulatory frameworks.

The ability of \acp{EV} to offer grid services is a well-researched topic, demonstrated by numerous studies. One of the earliest works, from 2005, showcased how bi-directional charging (V2G) could balance the utility grid by allowing \acp{EV} to charge and discharge based on grid conditions~\cite{Kempton2005}. Beyond reducing the burden on the grid through smart charging, \acp{EV} can complement the integration of renewable energy sources (RES) by storing excess energy during low-demand periods~\cite{Alkawsi2021}. This flexibility positions \acp{EV} as key enablers of the transition to renewable energy rather than passive energy consumers. Recent studies have further explored the role of \acp{EV} in mitigating grid challenges through implicit strategies~\cite{Sadeghianpourhamami2018QuantitiveApproach}. Several studies have quantified the flexibility potential of \acp{EV}, which can be harnessed through strategic pricing and operational strategies~\cite{Zade2020,Srensen2021}. For example, offering EV users incentives has alleviated localised grid issues in low-voltage (LV) distribution networks~\cite{Soares2022ElectricNetworks}. Time-of-use (TOU) rates have also proven effective for peak shaving via smart charging~\cite{BirkJones2022ImpactGrids}. However, TOU tariffs may require extensive consumer education, can be inaccessible to some users~\cite{2020EnergyTariffs} and even result in new charging peaks~\cite{panda2024aggregate}.

The previously mentioned studies highlight the numerous possibilities of utilising \ac{EV} flexibility, with success depending on factors such as incentive design, geographic and contextual conditions, regulatory frameworks, and the extent of \ac{EV} adoption and charging infrastructure development.  All of the above studies quantify the potential flexibility through different activation mechanisms, but they do not assess the dependability of that flexibility or how it scales across different levels of aggregation. However, the absence of an adequate regulatory framework leads to economic and institutional barriers, hindering the development of \ac{EV} flexibility services at the distribution network level. This contributes to increased uncertainty about the value of these services. For example, a range of European demonstrator projects analysing the adoption of \ac{EV} flexibility highlights the lack of regulatory support for using flexibility as a mainstream product~\cite{GonzalezVenegas2021ActiveServices}. While several researchers have demonstrated the technical feasibility of smart EV charging, there is limited literature assessing the dependability of quantified flexibility or how it scales across different levels of aggregation. In particular, very few studies quantify the risks associated with offering aggregate \ac{EV}  flexibility for mitigating congestion through market-based mechanisms. 

\subsubsection*{The Dutch case study}
Market-based congestion management services are currently being developed in several regions; however, this article focuses on the regulatory framework in the Netherlands. The Netherlands is at the forefront of addressing grid congestion challenges, driven by the rapid adoption of electric vehicles (\acp{EV}) and the simultaneous rollout of a dense public charging infrastructure. With over 180,000\footnote{\url{https://duurzamemobiliteit.databank.nl/mosaic/en-us/elektrisch-vervoer/laad--en-tankinfrastructuur-in-nederland}} public and semi-public charging points, the country has a charger-to-EV ratio of 0.1—one of the highest in Europe. This rapid growth has begun to strain the electricity grid, prompting both technical and regulatory responses.\par

To ensure that this growth in aggregate charging demand does not exacerbate grid congestion, the Dutch government has introduced forward-looking policies. The National Charging Infrastructure Agenda, part of the broader National Program for Charging Infrastructure, mandates smart charging functionality for all new \ac{EV} chargers. By the end of 2025, all chargers—private, semi-private, and public—must be internet-connected, capable of receiving external control signals, support secure communication via standardised protocols such as \ac{OCPI}, and be bi-directional charging ready~\cite{DutchBrochure, nal-wg-smart-charging-core-team-2021}. These technical requirements are designed to ensure that regulatory ambitions are matched with practical infrastructure readiness. Consequently, the Netherlands offers a unique and mature context in which to study the implementation and impact of market-based congestion management strategies.\par

Market-based capacity management can typically be achieved through services with reservation and activation components, where activation occurs if and when congestion is imminent, or through firm energy trades. In the Netherlands, the energy regulatory body has defined network congestion,  defined the characteristics of market products that can help alleviate congestion and assigned specific responsibilities to stakeholders participating in congestion management services~\cite{TheNetherlands2022DecreeManagement}. Details of these regulations have been elaborated in Section \ref{sec:2} \par

\subsubsection*{Contribution and organisation of the article}
The contributions of this paper are as follows:
In this article, we present how \ac{EV} charging stations can offer dependable congestion management products by quantifying their aggregate flexibility potential, using a real-life case study from the Netherlands. The presented analysis is conducted from the perspective of the \ac{CPO} or fleet operator; therefore, grid-side constraints such as voltages and currents are not considered.
\begin{itemize}
\item To support our analysis, we utilise a real-life data set from \ac{EV} charging stations in the Netherlands. It consists of charging stations in public and commercial spaces, and contains charging sessions from shared and privately owned vehicles. Charging stations are categorised based on typical charging behaviour (e.g., residential, shared, and commercial use), which is a key factor for understanding grid impacts and the flexibility of charging patterns.

 \item The statistical ability to deliver congestion management services at different aggregation levels and for various service types is analysed in detail. This analysis is carried out using a probabilistic model of charging behaviour, developed from real-world data.

\item We analyse the probability of meeting specific congestion management service requirements as a function of usage type and aggregation level. These probabilities will assist EV aggregators, such as charging point operators, in making informed decisions regarding offering congestion mitigation products. Additionally, they will help distribution system operators assess the potential of these products.

\item Finally, we demonstrate how machine learning models can effectively predict the capacity for offering flexibility products for congestion management services by charging stations. The impact of using such predictions for service provision is further explored. 
\end{itemize}
Finally, this article is organised as follows: Section~\ref{sec:1} introduces the growing adoption of \acp{EV} and its role in exacerbating congestion in power networks, while also motivating the use of \ac{EV} smart charging as a solution through market-based congestion management. The Netherlands is selected as a reference case due to its advanced regulatory framework and infrastructure readiness. Section~\ref{sec:2} delves into the technical details of congestion management products in line with Dutch regulations. Section~\ref{sec:3} outlines the key assumptions and describes the dataset used for the analysis. Section~\ref{sec:4} presents the modelling approach, including dispatch strategies and the use of machine learning models for forecasting flexibility. Section~\ref{sec:5} discusses the simulation results, and Section~\ref{sec:6} concludes the article with key insights and directions for future research.
\begin{table*}[t]
\caption{Comparison of redispatch and capacity limitation products for congestion management}
\label{tab:compa_product}
\begin{tabular}{lll}
\hline
\multicolumn{1}{c}{\textbf{Category}} & \multicolumn{1}{c}{\textbf{Redispatch}}             & \multicolumn{1}{c}{\textbf{Capacity Limitation}} \\ \hline
Purpose                               & Resolve post-market congestion                      & Prevent congestion via contracts                 \\
Timing                                & After day-ahead market                              & Pre-defined or day-ahead notice                  \\
Basis                                 & Market-based bids/offers                            & Contractual agreement                            \\
Minimum bid size  & minimum 100 kW                   & Varies by contract                               \\
Trigger                               & purchase of congestion-reducing transaction                   & Ttime-based or day-ahead notification                \\
Flexibility Needed                    & High                                                & Medium to High                                   \\
Cost/Benefit                          & Market revenue                                      & negotiated (fixed and/or activation fee)                    \\
Customer Role                         & free bids or compulsory bids ('biedplicht')               & bilateral agreement, sometimes compulsory                              \\ \hline
\end{tabular}
\end{table*}
\section{Congestion management regulations}\label{sec:2}
The functioning of a whole network (HV, MV or LV) can be compromised due to a single bottleneck, usually in the form of capacity limitations of cables, transformers or other equipment responsible for power transportation in the network. Alternatively, violation of voltage tolerances at specific nodes in the network can also pose significant challenges. Network congestion can occur due to electrification of heating, cooling, and the mobility sector, along with the growing penetration of distributed renewable energy sources~\cite{amscongestion}. 
In the Netherlands, electrification of demand and the rapid growth of distributed renewable generation have led to pervasive congestion in transmission and distribution grids, resulting in regular curtailment of generation and the delay of new connection requests. 
Measures to prevent and mitigate congestion in the Dutch power grid are being explored through alternative flexibility solutions to avoid the high costs and resource constraints of extensive network expansions in the short term. To encourage the use of flexibility as one of the mainstream solutions to avoid grid reinforcement, the Netherlands' energy regulatory body, the \ac{ACM}, has introduced regulations that clarify the concept of congestion, assign specific responsibilities to stakeholders participating in congestion management services, and defines the characteristics of market products that can help alleviate congestion~\cite{TheNetherlands2022DecreeManagement}.\par

Under the congestion management regulations, in the event of structural congestion, grid operators must inform all parties in the affected area about the nature, duration and magnitude of the expected congestion. Moreover, all available market-based resources must be used for congestion management before turning to non-market-based solutions (e.g., forced curtailment). The regulations introduce a new role in the Dutch grid code known as \acp{CSP}, responsible for offering flexibility to the grid operators. \acp{CSP} can place bids on behalf of connected parties with power-generating units, energy storage, or demand installations of 1 MW capacity or more. They can also participate in the redispatch process via counter-trading and act as aggregators of small-scale flexibility by submitting bids for a group of affiliates with power-generating units, energy storage, or demand installations under 1 MW capacity.\par

One of the key additions to these new regulations is that anyone with telemetry or a smart meter may offer flexibility services (through a \ac{CSP}). Still, those with a connection of over 60 MW are required to do so. Regulation exemptions may be made for national security, hospitals, and other entities. In congested areas, smaller consumers with a connection capacity of greater than 1 MW and less than 60 MW may be required by the \acp{DSO} to participate in congestion services. 

Market-based congestion management in the Netherlands is based on two types of products with various subtypes:
\begin{itemize}
    \item \textit{Redispatch}: Adjustments relative to the scheduled consumption after the closure of the day-ahead market. This is achieved through matching intra-day bids and offers inside and outside the congested region, mediated by the GOPACS platform.\footnote{\url{https://www.gopacs.eu}} To participate in the congestion redispatch market, a \ac{CSP} must offer a minimum of 100 kW for one imbalance settlement period. Two variations exist:
    \begin{itemize}
        \item Mandatory participation (\emph{biedplichtcontract}). Large customers must offer redispatch products, often for a predetermined price. 
        \item Voluntary participation is enabled by trading on the intraday market with the inclusion of a location tag that enables matching trades across congestion boundaries. 
    \end{itemize}
    \item \textit{Capacity limitation:} Standard connection agreements in the Netherlands have a fixed capacity. Various options have been (and are being) created to offer more flexible connection contracts. 
    \begin{itemize}
        \item Fixed capacity limitation contracts. For a fee, customers agree to reduce their contracted capacity at specified times.
        \item On-demand capacity limitation contracts. As mentioned above, the capacity reduction only applies when the network operator sends a notification through the GOPACS platform. The notification is sent on the previous day's morning so that it can be accounted for in day-ahead trading.
        \item Non-firm connection agreement. Connections are offered at a reduced cost but come with limitations on usable capacity. These limitations may be static or dynamic, depending on the contract type. 
        \footnote{Although formally not a market-based congestion management product, the non-firm connection agreement is included in this listing because, on operational time scales, it works the same way as the capacity limitation contracts~\cite{non_stiff_code}}
    \end{itemize}
\end{itemize}
In this article, we consider two main types: redispatch and capacity limitation, which are summarised in Table \ref{tab:compa_product}.

\section{Simulation Setup and Data Overview}\label{sec:3}
For the rest of the paper, the following definitions and arrangements hold. The charging topology used in this paper consists of three entities:
\begin{itemize}
    \item \textit{Charging point (CP)} refers to the physical connector as a socket or a cable connection that an \ac{EV} can use to charge.  In \ac{OCPI}, this is referred to as an \ac{EVSE}, which is responsible for the power supply to a single \ac{EV} in a single session.
    \item \textit{Charging station (CS)} refers to a charging pole containing one or more \acp{CP}. The typical arrangement for charging points in public or commercial settings (used in this study) is a CS with two \acp{CP}.
    \item \textit{Charging session}: Refers to the session envelope that is determined by the connection and disconnection times, energy requirements, and power capabilities associated with charging an \ac{EV} at one of the \acp{CP} within a \ac{CS}.
\end{itemize}

\par

\subsection{Dataset}
The analysis presented in this paper is based on over 550,000 \ac{EV} charging transactions that occurred between June 2020 and March 2023. All the charging events are associated with \acp{CS} operated by a single \ac{CPO} distributed in and around Utrecht, The Netherlands. The data is extracted using the \ac{OCPI} protocol, an open-source, scalable protocol that enables a seamless connection between a \acp{CPO} and any e-mobility service provider~\cite{Ocpi_git}. The time-series power profile for all charge points was extracted first and then linked with each charging session based on the starting and ending times of connections. The maximum charging power ability of \acp{EV} in each charging session was obtained by finding the maximum recorded power of its associated charging profile.\par

\begin{figure*}[t]
    \centering
    \includegraphics[width=1\linewidth]{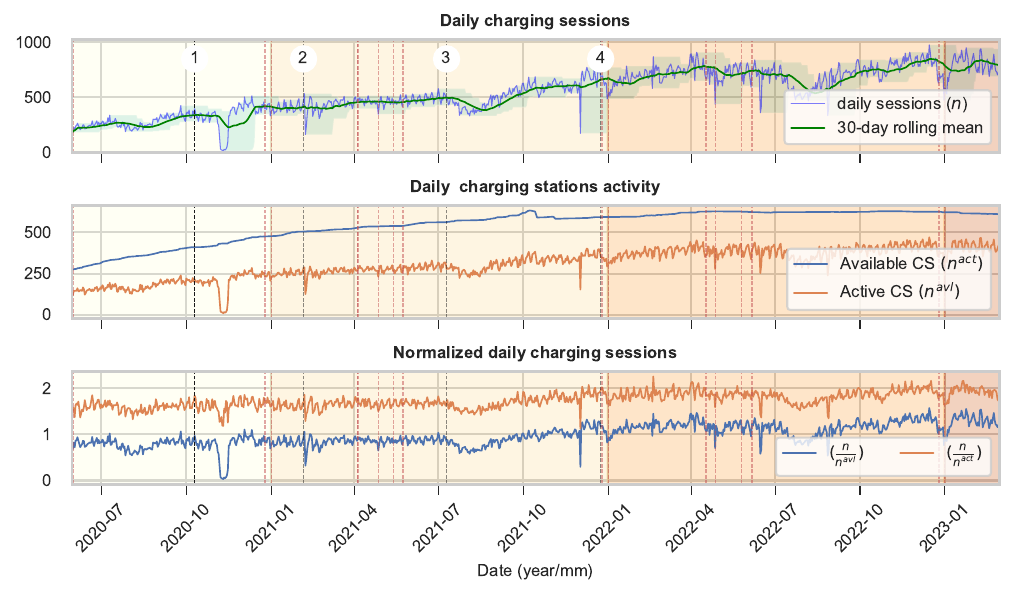}
    \caption{Evolution of charging events over time. The top figure shows the trends in daily charging transactions, while the middle plot shows the evolution of daily available and active \acp{CS}. The bottom plot shows the ratio of daily charging events to that of daily active and available \acp{CS}. Data spans from May 2020 to February 2023. Measures related to the coronavirus pandemic are shown in \blackDashedLine{}, where annotations 1, 2, 3, and 4 reflect major events, such as the start of partial lockdown, begin of night curfews, the closure of catering/ social places and the start of the second lockdown respectively in the Netherlands. Background shading differentiates various years. Also, public holidays that might affect charging patterns are shown using \redDashedLine{}.The dip in transactions observed in November 2020 is due to a data error caused by the measurement device being offline at the source.}
    \label{fig:tra_growth}
\end{figure*}
\subsection{Statistical insights}
There has been a growth in the \ac{EV} charging infrastructure and the number of daily charging sessions, as seen in \figref{fig:tra_growth}; however, this growth has been uneven due to various factors. For example, sharp variations in daily charging events could be seen on days near the introduction/ withdrawal of certain coronavirus-related restrictions. The large-scale effect of the pandemic on the \ac{EV} industry has been analysed in detail in ~\cite{Wen2021ImpactsChina}. Periodic events such as public holidays also impact the \ac{EV}'s usage pattern, which in turn affects the daily charging transactions as shown in \figref{fig:tra_growth}.\par

The number of daily \ac{EV} charging events is strongly correlated with the increase in \acp{CS}. We distinguish daily \emph{available} \acp{CS},  which were available on a given day, whether or not they were utilised by any \acp{EV}, and daily \emph{active} \acp{CS},  which have been utilised at least once during the day for \ac{EV} charging. By definition, the number of daily active \acp{CS} is always less than or equal to the number of daily available \acp{CS} for a given day. 
\par

As seen in \figref{fig:tra_growth}, there has been substantial growth in daily charging events over the past years. When the number of charging sessions is normalised by the number of daily active or available \acp{CS}, we observe a relatively steady number of charging events per available/active \ac{CS} across all three years of data. Among those, the ratio of charging events to daily available \acp{CS} shows less fluctuation. For this reason, the stochastic model introduced in later sections samples charging sessions by drawing sets of sessions associated with available \acp{CS} on a random day. 

\subsection{Charging locations and accessibility}

\begin{figure}
    \centering
    \includegraphics[width=0.8\linewidth]{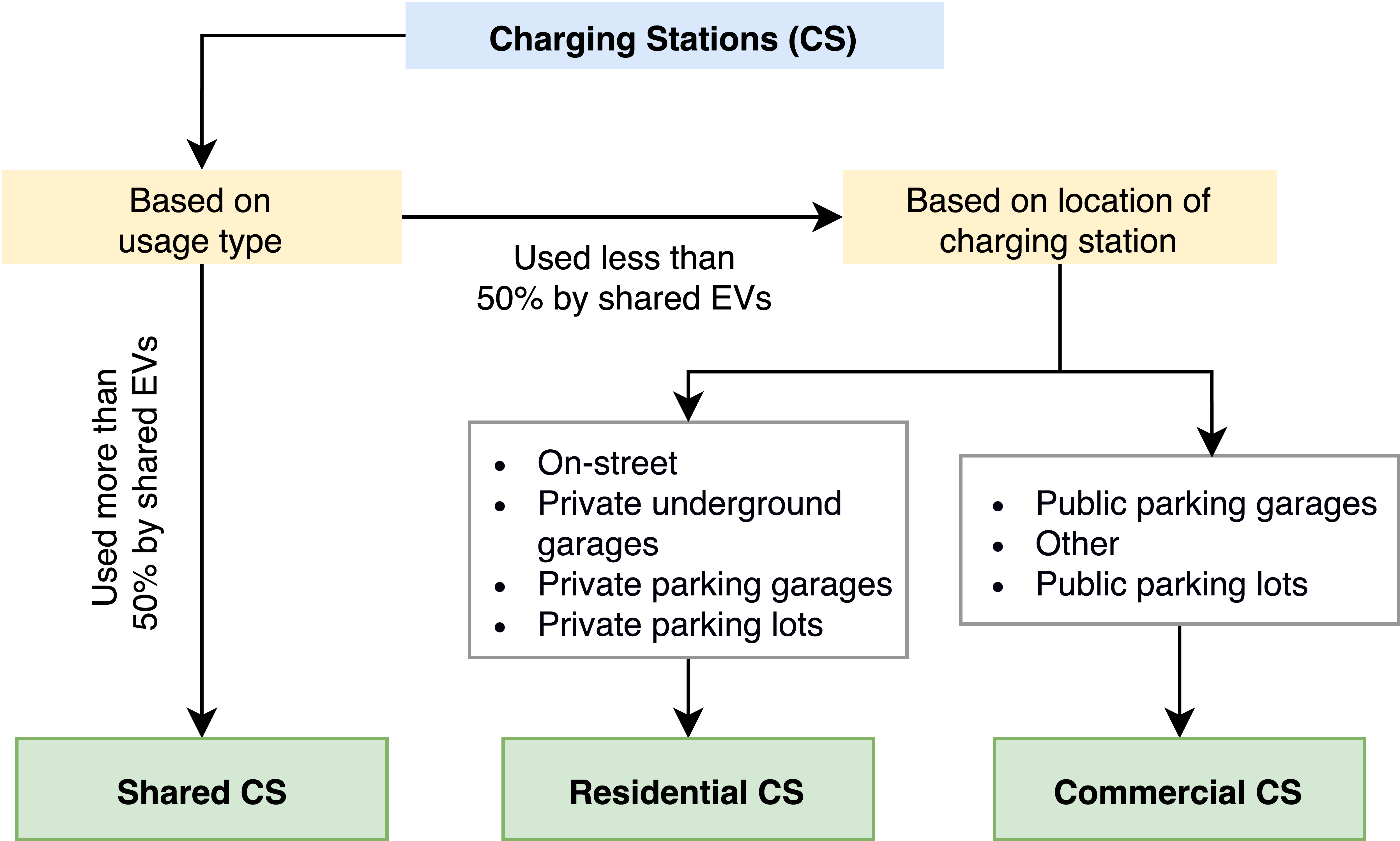}
    \caption{Method of categorisation for \acp{CS} based on location, accessibility and type of usage.}
    \label{fig:cat_process}
\end{figure}

Although \acp{EV} charging schedules are highly stochastic, their properties depend on the time of the day, location of charging and category of \ac{EV} users. This is most easily explored through average or aggregate charging profiles. After a visual exploratory analysis, we categorised all the available \acp{CS} into three categories and labelled \emph{shared}, \emph{commercial} or \emph{residential} based on their characteristics as shown in \figref{fig:cat_process}. Shared \acp{CS} are predominantly used by shared \acp{EV} (more than 50\% sessions). In contrast, residential and commercial \acp{CS} are categorised based on location, as locations influence the charging patterns of the \acp{EV} charged there.\par

In the analysed dataset, all the charging sessions are from \acp{CS} operated by a single \ac{CPO}. These \acp{CS} are located at different places, some with accessibility restrictions. Using the accessibility tags and the type of location, we segregate the charging stations into residential or commercial. For instance, on-street chargers that are publicly accessible are considered residential because they can be used by anyone who owns an \ac{EV}. Also, the charging stations located at private locations have limited accessibility due to their physical location, such as being inside a gated property. This means that only individuals with access to the private location, such as residents or employees, can use the \acp{CS} located there. The categorisation of charging sessions based on location and accessibility is depicted in \tableme{tab:division_loctype}.\par

\begin{table}
\centering
\caption{Distribution of charging sessions based on charging location and accessibility.}
\label{tab:division_loctype}
\begin{tabular}{lcc} 
\toprule
\% of sessions     & Public                                  & Private                                  \\ 
\midrule
On street          & {\cellcolor[rgb]{1,0.855,0.592}}78.6    & {\cellcolor[rgb]{1,0.855,0.592}}1.8      \\
Unknown            & {\cellcolor[rgb]{1,0.855,0.592}}1.8     & {\cellcolor[rgb]{1,0.855,0.592}}8.7      \\
Other              & {\cellcolor[rgb]{0.741,0.741,0.945}}0   & {\cellcolor[rgb]{0.741,0.741,0.945}}2.0  \\
Parking lot        & {\cellcolor[rgb]{0.741,0.741,0.945}}3.3 & {\cellcolor[rgb]{1,0.855,0.592}}1.7      \\
Parking garage     & {\cellcolor[rgb]{0.741,0.741,0.945}}1.7 & {\cellcolor[rgb]{1,0.855,0.592}}0.3\\
Underground garage & {\cellcolor[rgb]{1,0.855,0.592}}0       & {\cellcolor[rgb]{1,0.855,0.592}}0.1      \\ 
\hline
\multicolumn{3}{l}{
\begin{tabular}{l l l l}
        {\cellcolor[rgb]{1,0.855,0.592}} & Residential & {\cellcolor[rgb]{0.741,0.741,0.945}} & Commercial
\end{tabular}}                                                                                    \\
\bottomrule
\end{tabular}
\end{table}

\begin{figure}
    \centering
    \includegraphics[width=\linewidth]{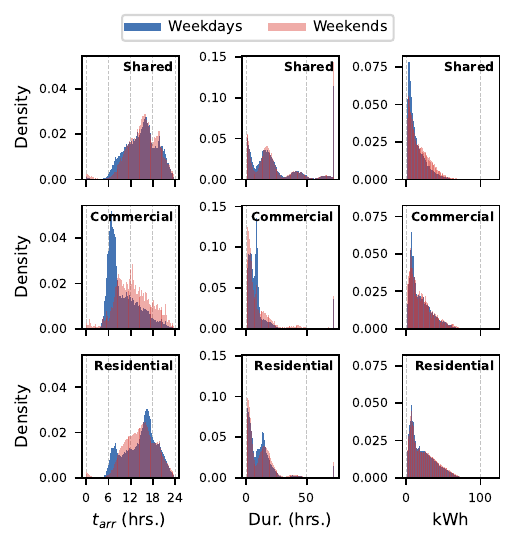}
    \caption{Distribution of arrival time, parking duration and charging energy demand for the charging sessions in each category of 
    \acp{CS}.}
    \label{fig:cat_prop}
\end{figure}

The distinct charging behaviour in these three categories is evidenced by the statistical distributions shown in \figref{fig:cat_prop}. Shared \acp{EV} transactions have a similar distribution of arrival times with transactions from residential \acp{CS} during weekdays, except for fewer new arrivals in the morning hours. However, transactions in commercial locations experience a morning peak in activity when people arrive at work and connect their vehicles to charging stations. The evening peak for residential  \acp{CS} is attributed to vehicle connections made after returning from work. On the other hand, shared \acp{CS} are primarily occupied during evening hours as the shared \acp{EV} is used in the morning and gradually returned during the evening hours.\par

Based on the distribution of connection times, it can be inferred that \acp{EV} are connected for significantly more extended periods in shared \acp{CS} compared to \acp{EV} in residential and commercial \acp{CS}. This trend may be due to lower reservation rates for shared \acp{EV} during weekends compared to weekdays. Analysis of the distribution of charging volume requirements also shows that \acp{EV} using shared \ac{CS} exhibit greater homogeneity in terms of charging volume compared to other \ac{CS} categories.

Arrival times and connection duration differences between weekdays and weekends for all three categories of \ac{CS}. During the weekend, the arrival time peak for \acp{EV} in shared \ac{CS} is shifted back a few hours, possibly due to different usage patterns during weekdays and weekends and varying charging infrastructure availability. We also observe a group of very late arrivals after midnight, likely consisting of \acp{EV} used for more extended leisure or sightseeing trips. For commercial \acp{CS}, there is a reduction in the morning peak on weekends attributed to the closure of work locations. Most weekend arrivals at commercial charging poles are at places that can also be used by the public, such as shopping complexes or tourist attractions. For residential users, the morning and evening peaks are flatter. We see considerable peaks in connection duration for shared transactions over the weeks. This is mainly due to low reservation rates of shared \acp{EV} during the start of the week. For residential charging sessions, the connection duration shows less variability due to the prevalence of occupancy fees (parking for an extended duration) in most of the on-street charging locations.

\begin{figure*}
    \centering

\includegraphics[width=\linewidth]{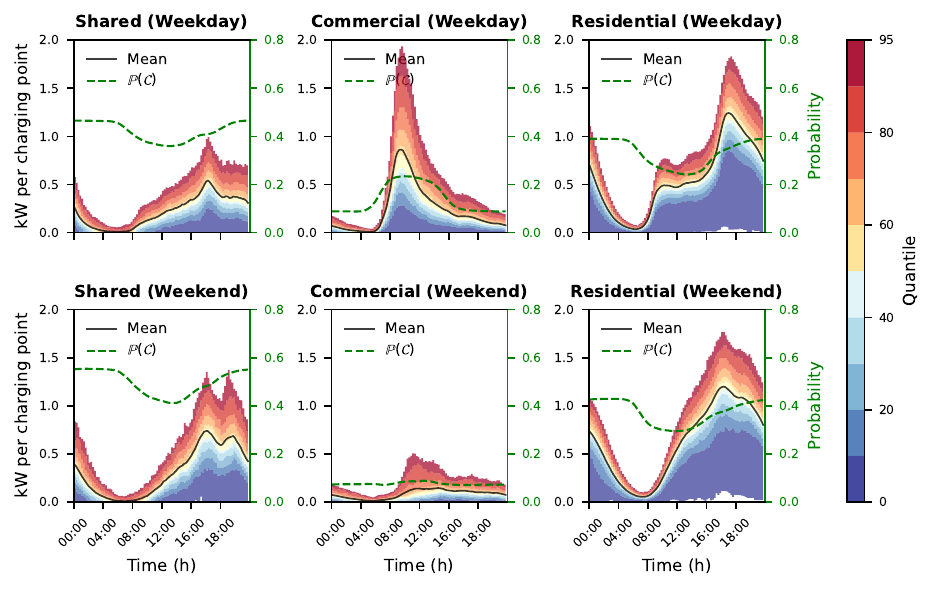}
\caption{Quantile distribution of unoptimised aggregate load profiles (normalised by the number of available charging points), segregated by \ac{CS} category and weekday/weekend. The empirical probability of an \ac{EV} being connected to one of the \acp{CP} is also plotted using a secondary axis. Charging profiles are computed based on the arrival time, energy transfer and maximum charging power associated with the charging sessions for each category.}

    \label{fig:load_quantile}
\end{figure*}

The differences among the three categories of \acp{CS} can also be seen in their aggregated daily unoptimised charging profiles as shown in \figref{fig:load_quantile}, separated by day of the week (weekday or weekend). Unoptimised charging refers to the charging strategy where each \ac{EV} is charged with the maximum possible power as soon as it is connected to the charger until it is fully charged. The data underlines the difference in charging behaviour for the three identified categories and between weekdays and weekends.

\figref{fig:load_quantile} also shows the weighted average 15-minute probability ($\mathbb{P}_t(C)$) of an EV being connected to one of the \acp{CP} within a specific CS category ($C$). To calculate this probability, we first determine the probability of an \ac{EV} being connected per quarter hour for each \ac{CP}:
\begin{equation}
    \mathbb{P}_t(p) = \frac{\sum_{d\in \mathcal{D}_{p}}a_{p,d,t}}{|\mathcal{D}_{p}|}, \quad\forall p\in C, \: \forall t \in \{1,\dots,24\times 4\}. \label{prob1}
\end{equation}
Here, $\mathcal{D}_{p}$ refers to the collection of continuous days during which a \ac{CP} is operational (available). The binary variable $a_{p,d,t}$ indicates whether a connection between an \ac{EV} and a \ac{CP} exists at a specific 15-minute interval  $t$ on a given day $d$.
Subsequently, we compute the weighted average of these probabilities across all \ac{CP} within the same category, as explained in \eqref{prob2}. The weights ($w_{p}$) correspond to the ratio of the operational time (measured in days) of each \ac{CP} to the total time span (also in days) of the dataset.
\begin{equation}
    \mathbb{P}_t(C) = \frac{\sum_{p\in C} w_{p}\mathbb{P}_t(p)}{\sum_{p \in C} w_{p}}, \quad\forall t \in \{1,\dots,24\times 4\}.  \label{prob2}
\end{equation}

\section{Models and Methods}\label{sec:4}
\begin{figure*}
\centering
\fbox{\begin{minipage}[b]{0.99\textwidth}
\begin{multicols}{2}
\printnomenclature
\end{multicols}
\end{minipage}}
\end{figure*}
For each of the \ac{CS} categories, a statistical model is constructed using historical charging transactions. This model is subsequently utilised to generate random input samples for various dispatch strategies, thereby enabling an evaluation of their effectiveness in delivering a specific flexibility product. First, we explain how sampling is done to analyse historical charging transactions.

\subsection{Random sampling of charging events}
Initially, our analysis aims to assess the probability of offering various market products (redispatch and capacity limitation) through the \acp{CS} of various pre-determined categories using the set of historical charging transaction data. For this, we perform a Monte Carlo analysis using deterministic dispatch models with different objectives for each market product. These models are applied to randomly selected charging events, sampled from historical data using the process explained in Algorithm \ref{mc_dispatch_alg}. Sampling is performed by drawing a given number of historical traces from the same day, for a given \ac{CS} category, ensuring that the quantity of daily available \acp{CS} used for sampling does not surpass the lowest daily available \acp{CS} for that category in the data.\par

The Monte Carlo samples generated are then utilised to estimate the likelihood of offering a particular type of flexibility product. The characteristics of these products are summarised in Table \ref{tab:compa_product}. Under the congestion management products outlined in Section \ref{sec:2}, we focus on two flexibility products: redispatch and capacity limitations.\par

\begin{algorithm}[!ht]
\DontPrintSemicolon
      \KwInput{
      \begin{itemize}
      \item $\mathcal{D}$: Set of unique dates available in the charging transaction data
      \item $\mathcal{C}$: Set of \ac{CS} categories (shared, residential \& commercial)
      \item $\mathcal{Q}$: Set of quantities of \ac{CS}, for which simulation is run
      \item $n_{sample}$: number of sampled days
          \item $\mathcal{D}_{CS}(d,c)$: Set of daily available \acp{CS} on $d\in\mathcal{D}$ for category $c\in \mathcal{C}$
      \end{itemize}}
\KwOutput{Monte Carlo iteration results for the desired model.}
\For{$c\in \mathcal{C}$}
    {  
            \For{$q\in \mathcal{Q}$}    
                {
                    initialize: $n$ = 0 \\
                    \While{ $n< n_{sample}$}
                    { 
                        $d_{s}$:= random sampled date from $\mathcal{D}$ \\
                        \If{n $\geq$ $|\mathcal{D}_{CS}(d_s,c)|$ }
                        {
                            escape from while loop
                        
                        }
                        \Else
                        {Select q random CS from the set\\
                        Solve the desired model, analyse and store the results\\
                        $n$ += 1
                    
                        }
                    }  
                }    
        
    }
\caption{Pseudocode for Monte Carlo analysis.}
\label{mc_dispatch_alg}
\end{algorithm}
To analyse the above models, we compare results under different dispatch strategies, as explained in the following section. 
\subsection{Dispatch Strategies}\label{dispatch_stratgies_section}
In this paper, we use four dispatch strategies for each of the \ac{CS} categories, including unoptimised 
charging of \acp{EV}, to compare their effectiveness in delivering different market products. Each of these strategies is based on the following assumptions:
\begin{itemize}
    \item All \acp{EV} are charged fully by the end of their session. The volume of energy each vehicle charges is taken from the historical data, which can be less than or equal to the capacity of the \ac{EV}'s battery.
    \item Apart from charging fully, for the bi-directional charging of \acp{EV}, the \ac{EV}  can only start discharging after charging first, so that at no given point the \ac{SOC} of the batteries is less than their initial \ac{SOC}. This assumption stems from the fact that the total battery capacity is not available in the session information, so it is unclear whether lower \acp{SOC} values are accessible. This also ensures that an \ac{EV} owner never finds their vehicle with a lower \ac{SOC} than they initially had. For the purpose of this paper, the impact of discharging on the \ac{EV}'s battery is not considered, as it is out of scope.
    \item \acp{EV} are not allowed to charge more than their corresponding historical charge volume, which potentially limits the actual flex capability of \acp{EV} and results in a conservative estimation of flexibility. 
    \item The time resolution of available data is in seconds. However, it was resampled to a time resolution of 15 minutes, which is also the time step used for all the analyses. The arrival and departure times are rounded to the nearest 15 minutes. Transactions that became infeasible after rounding due to having a connection time lower than the minimum time required for charging were dropped (fewer than 0.1\% of transactions)
    
    \item The maximum duration for which an \ac{EV} can be connected to a charger is restricted to 36 hours (144 time steps).
\end{itemize}
\begin{figure*}
    \centering
    \includegraphics[width=\linewidth]{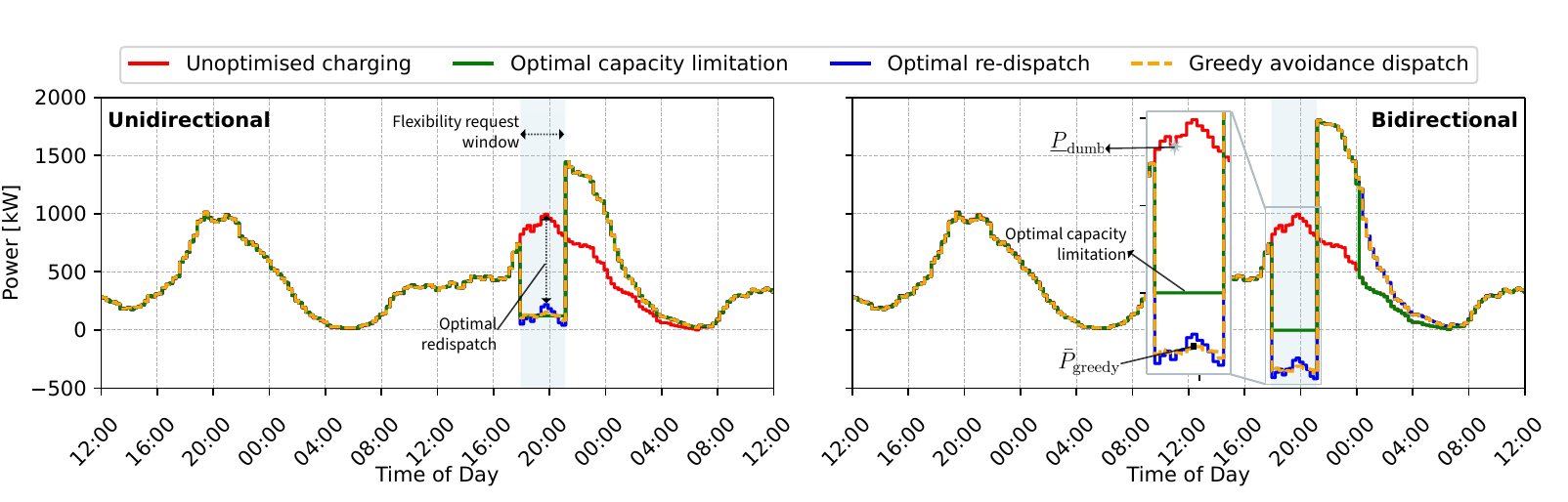}
    \caption{Aggregated power profiles for different dispatch strategies as explained in Section \ref{dispatch_stratgies_section} for a randomly selected day with 250 residential type \acp{CS} sampled from all available \acp{CS} on that day used predominantly by non-shared residential \acp{EV} users.}
    \label{fig:dispatch_startegies}
\end{figure*}
\subsubsection{Unoptimised charging}\label{unopt_model_formulation}
\emph{Unoptimised} charging ( \emph{dumb}/ as fast as possible) is conventionally used without other control strategies (smart charging). The \acp{EV} are charged using maximum power until their original charging demand ($\overline{e}_n$) is met.

The optimal dispatch strategies assume perfect knowledge of EV arrival and departure times. While this assumption can lead to overly optimistic flexibility estimates, it may still be realistic in cases such as shared charging stations, where reservations for shared vehicles are known in advance.

To provide a more realistic benchmark, we also consider a greedy dispatch strategy, where EV arrivals are not known beforehand. However, departure times are assumed to be known, which could be feasible in practical settings—either inferred from historical charging patterns or provided manually by the user upon plugging in.
\subsubsection{Optimal redispatch}\label{redispatch}
The optimal redispatch strategy is based on the assumption of having perfect knowledge of the arrival and departure times of all \acp{EV} to maximise the deviation between the optimised aggregated power profiles and the \ac{BAU} scenario. This paper considers the \ac{BAU} aggregate profile as the unoptimised case (ref. Section \ref{unopt_model_formulation}). The deviation is maximised for all time steps within the flexibility request time window ($\mathcal{T}_f$) as shown in \figref{fig:dispatch_startegies}. The objective is to maximise $c^r$, representing the shift from the \ac{BAU} aggregate profile inside $\mathcal{T}_f$.

The optimal redispatch capacity ($c^r$) is calculated by solving a simple linear program defined by the set of equations \eqref{eq:obj1}-\eqref{eq:all_6}. A small multiplier $\epsilon$ is used for the second term to ensure unique solutions. It selects those that charge the vehicles most rapidly within the set of solutions with the same (optimal) redispatch capacity. 
Equations \eqref{eq:all_1}-\eqref{eq:all_3} show the relationship between power and energy and ensure that all \acp{EV} are fully charged when they depart from the charging station. The charging and discharging power is limited through the constraints \eqref{eq:all_4}-\eqref{eq:all_5}. The optimal redispatch policy is enforced by constraint \eqref{eq:all_6} and the objective function \eqref{eq:obj1}.
\begin{equation}
    \max_{c^r, p, e}\: c^r + \epsilon \left( \sum_{n\in \mathcal{N}}\sum_{t\in \mathcal{T}}e_{n,t}\right)
    \label{eq:obj1}
\end{equation}
subject to:
\begin{align}
    &e_{n,t} = 0, \span t\leq t_n^a \:\: n\in\mathcal{N}\label{eq:all_1}  \\
    &e_{n,t} = e_{n, t-1}+p_{n,t-1}\Delta t, \:\:\span  t_n^a <t<t_n^d\:\: n\in\mathcal{N}\label{eq:all_2}\\
    &e_{n,t} = \overline{e}_n, \span t\geq t_n^d  \:\: n\in\mathcal{N}\label{eq:all_3}\\
    & p_{n,t}  = 0, \span t<t_n^a \wedge t\geq t_n^d \:\: n\in\mathcal{N}\label{eq:all_4}\\
    & \underline{p}_{n}\leq  p_{n,t} \leq \overline{p}_{n} \span t_n^a\leq t < t_n^d \:\: n\in\mathcal{N}\label{eq:all_5} \\ 
&\sum_{n\in\mathcal{N}}p_{n,t} \leq P_t^{base} - c^r, \span t\in\mathcal{T}_f \label{eq:all_6}
\end{align}

\subsubsection{Optimal capacity limitation}
Optimal capacity limitation strategies dispatch \acp{EV} in such a way that minimises the maximum aggregate power during a given flexibility request window. Like the optimal redispatch policy, this also assumes perfect knowledge of all \ac{EV}'s arrival and departure times. A linear program-based optimisation can minimise the value $c^l$ in \eqref{eq:optimal_cap}. The optimisation formulation for capacity limitation is similar to that of redispatch, except for the sense of optimisation and the constraint  $c^l$ as shown in \eqref{eq:cap_lim_con}. As $c^l$ is considered positive, the effect of uni- or bidirectional charging on capacity limitations will be the same.\par
The optimal peak capacity required during the flexibility request window can be leveraged for non-firm capacity contracts by the \acp{CPO}. For a single case, the calculated capacity limitation is shown in \figref{fig:dispatch_startegies}.
\begin{align}
    \min_{c^l, p, e}\quad c^l - \epsilon \left( \sum_{n\in \mathcal{N}}\sum_{t\in \mathcal{T}}e_{n,t}\right)\quad\label{eq:optimal_cap}
\end{align}
subject to:
\begin{align}
    &\sum_{n\in\mathcal{N}}p_{n,t} \leq c^l , \span  t\in\mathcal{T}_f \label{eq:cap_lim_con}\\
    &\text{Constraints } \eqref{eq:all_1}-\eqref{eq:all_5}\nonumber
\end{align}

\subsubsection{Greedy congestion avoidance dispatch}
The greedy dispatch algorithm schedules EV charging sessions so each session \emph{individually} minimises its power consumption during congestion. This results in an aggregate contribution smaller than for both optimal strategies (both consider the interdependencies between sessions), but the greedy strategy requires no knowledge of future EV arrival times. Implementing this strategy is sufficient if one knows the departure time and energy requirement at the start of the session. This is practically conceivable when using, e.g., in-car preferences, app-based control, or an interface on the charging station. 

Similar to other dispatch policies, the greedy dispatch can be realised by minimising the charging power of each \ac{EV} within the flexibility request window $\mathcal{T}_f$ using the objective \eqref{eq:greedy} along with the following linear constraints. 
\begin{align}
    \min_{c^l, e}\sum_{n\in \mathcal{N}}c^l_{n}- \epsilon \left( \sum_{n\in \mathcal{N}}\sum_{t\in \mathcal{T}}e_{n,t}\right)\label{eq:greedy}
\end{align}
subject to:
\begin{align}
    p_{n,t} &\leq  c^l_n, && t\in\mathcal{T}_f \label{eq:cap_lim_greedy1}\\
    c^l_n &\geq 0, && n\in \mathcal{N}  \label{eq:cap_lim_greedy2} \\
    \text{Constraints } &\eqref{eq:all_1}-\eqref{eq:all_5} \nonumber
\end{align}
$c^l_n$ is the optimal capacity limitation per vehicle. The aggregate capacity limitation of is given by $\max\left(\sum_{n\in \mathcal{N}}p_{n,t}^*\right)$, $t\in \mathcal{T}_f$, where $p_{n,t}^*$ is the optimised individual power profile. Similarly, the greedy redispatch can be calculated with respect to the base profile (unoptimised) as $\min\left(P_t^{base}-\sum_{n\in \mathcal{N}}p_{n,t}^*\right)$, $t\in \mathcal{T}_f$.

Figure \ref{fig:dispatch_startegies} shows the aggregated power profile for different mono- and bi-directional charging dispatch strategies. Although bi-directional charging performs much better in the case of a redispatch strategy, it does not influence flexibility potential in the capacity limitation dispatch policy, as the lower bound of capacity limitation is 0.
\begin{figure}
    \centering
    \includegraphics[width=\linewidth]{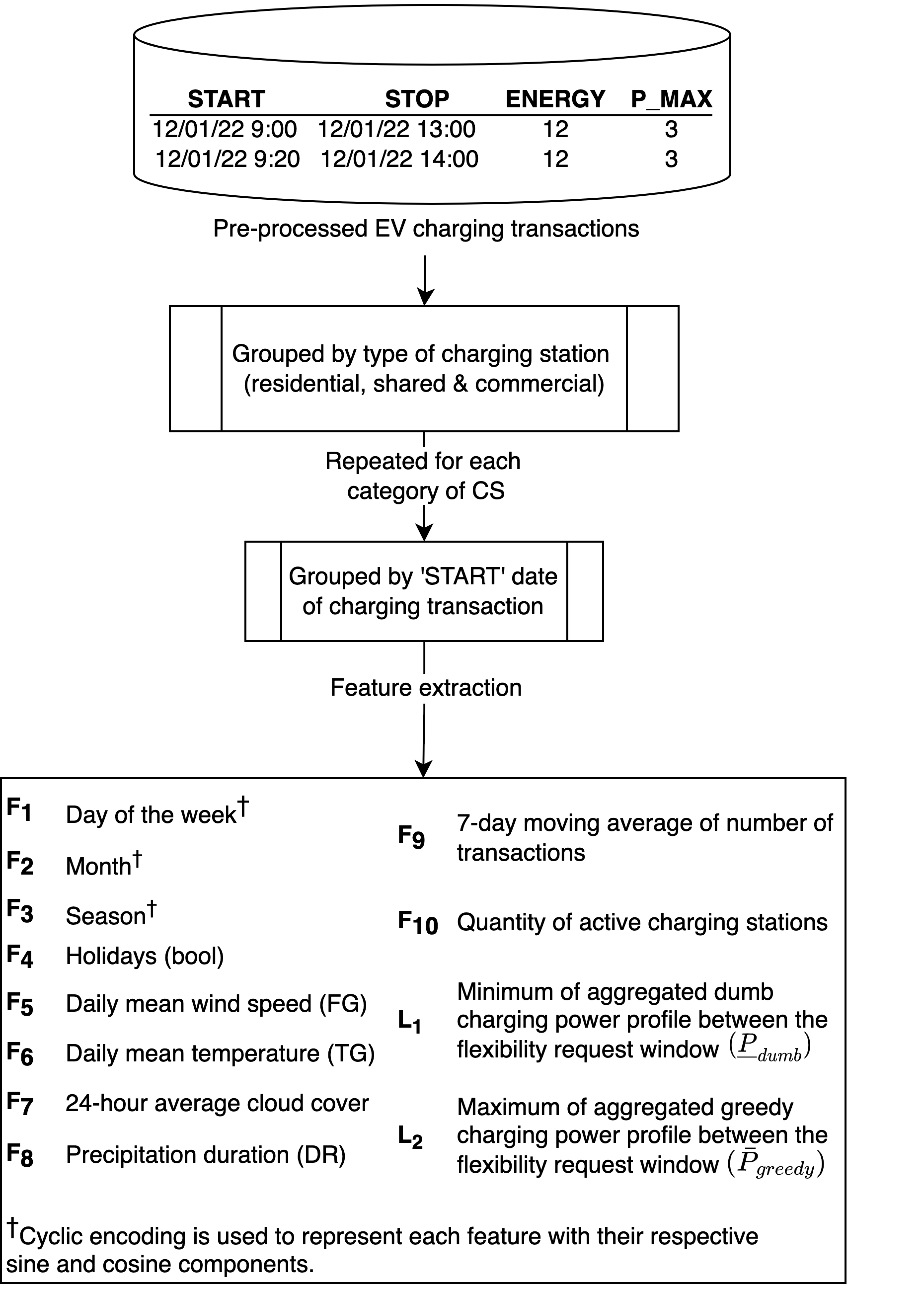}
    \caption{Workflow showing extraction of features from pre-processed \ac{EV} transactions data.}
    \label{fig:ml_workflow_data}
\end{figure}
\begin{figure*}
    \centering
    \includegraphics[width=1\linewidth]{updated_figures/ml_workflow_latest.png}
    \caption{Train and test workflows for the proposed machine learning models $\mathcal{M}_1^{da}$ and  $\mathcal{M}_2^{rd}$}
    \label{fig:ml_workflow_tr_test}
\end{figure*}
\subsection{Forecasting of flexibility}\label{ML_work_flow_sections}
The models outlined in Section \ref{dispatch_stratgies_section} rely on having perfect foresight to determine optimal values for delivering various market products. However, real-world scenarios are characterised by inherent uncertainty regarding the departure and arrival times of each \ac{EV}. For example, a study focused on EV users in the Netherlands showed that the chosen route by drivers can impact the \ac{SOC} of each vehicle upon arrival at the charging location and the availability of different types of chargers, such as fast or slow, can influence the EV user's route selection~\cite{Ashkrof2020AnalysisNetherlands}. As a result, a single charging session can be influenced by numerous uncertainties, which in turn can ultimately affect the \ac{EV} fleet's flexibility. This uncertainty can impact the overall aggregated charging profile, subsequently affecting the feasibility of delivering redispatch flexibility products. \par
We use machine learning models to address this inherent uncertainty and understand its impact on the probability of providing redispatch products. We consider a setting in which a \ac{CPO} makes a day-ahead prediction of the anticipated inflexible demand and then offers congestion management services relative to that anticipated load level. In addition, the \ac{CPO} then forecasts the magnitude of the service provision. \par

\subsubsection{Feature engineering}
The first step involves curating a set of features necessary for getting accurate predictions. The features incorporated in our approach are derived from key factors that have the potential to influence the usage patterns of \acp{EV}. Four date-related features are the day of the week, season, month and holidays. Days of the week and seasons are transformed into circular coordinates using trigonometric transformations, whereas the holiday feature is boolean. These features and the 7-day moving average are known by \ac{CPO}. In addition, several weather-related features are chosen due to their influence on the usage and driving pattern of \acp{EV}~\cite{Donkers2020InfluencePerformance}. For example, a less sunny and windy day will attract more users to use their cars rather than their bicycles.  How seasonal variations can affect \ac{EV} usage among different categories like private and shared \acp{EV} is explained in detail in \cite{Hao2020SeasonalVehicles}. The daily mean temperature, sunshine duration, daily mean wind speed and precipitation duration are weather-related features. \par

We propose two models: $\mathcal{M}_1^{da}$ and $\mathcal{M}_2^{rd}$, as explained, which use the designed features to forecast day-ahead base profile and redispatch power, respectively.

\subsubsection{Forecasting day-ahead profile ($\mathcal{M}_1^{da}$)}
For predicting redispatch power within the flexibility request window, estimating the \ac{BAU} power profile first is crucial, as it influences how much the \ac{EV} fleet can deviate. We begin by forecasting the day-ahead base load profile (unoptimised charging). Relative to this forecast, we calculate the redispatch potential achievable when individual vehicles are dispatched using a greedy congestion avoidance dispatch strategy. This method inherently considers the uncertainties present in the base profile, as this is known on a day-ahead basis and is consistent with the information available to aggregators when formulating their redispatch bids.

\subsubsection{Forecasting redispatch power ($\mathcal{M}_2^{rd}$)}
Subsequently, we train a second machine learning model to forecast the redispatch power directly. This model leverages the output from the first model (i.e., the redispatch derived from the greedy method based on the output of $\mathcal{M}_1^{da}$) as its training target. Once trained, the aggregator can use the second model operationally to estimate redispatch potential in cases where the \ac{CPO} must also announce its redispatch ability before actual realisation. This methodology enables us to account for uncertainties in arrival/departure times and energy requirements in combination with a simple dispatch strategy (assuming only that energy and departure times are known at the start of the charging session). 

\subsubsection{Training and testing}
As shown in \figref{fig:ml_workflow_data}, the pre-processed charging transactions are grouped per their start date and sampled based on Algorithm \ref{mc_dispatch_alg}. Each sampled set is represented by its features $F_{\#}$ for the training of the predictive models (c.f. \figref{fig:ml_workflow_data}).\par

First, a \ac{SVR} model ($\mathcal{M}_1^{da}$) is trained using the set of features $X=\{F_1, \ldots, F_{10} \}$  to predict $L_1 \equiv \underline{\hat{\text{P}}}_{dumb}$ (c.f. \figref{fig:ml_workflow_tr_test}), which is the minimum of the unoptimised aggregate charging profile within the flexibility request window. Next, the trained model is used to create the new target labels: $\Delta L = \hat{L}_1-L_2$, which is a conservative estimate of redispatch power from the aggregate unoptimised \ac{BAU} profile, where individual \acp{EV} are dispatched per the greedy congestion avoidance dispatch algorithm. Next, the same set of features, but with the new label, is used to train $\mathcal{M}_2^{rd}$ (also using SVR), which predicts the redispatch power ($\hat{\text{P}}_{redispatch}$). 
The pair of models can be summarised as
\begin{align}
    \hat{L}_{1}  &= \mathcal{M}_1^{da}(X) \label{ml_1},&& [\textrm{predictor of }L_1]\\
     \widehat{\Delta L} &= \mathcal{M}_2^{rd}(X), && [\textrm{predictor of }\hat{L}_1-L_2] \label{ml_2}
\end{align}
The entire training and testing workflow is visualised in \figref{fig:ml_workflow_tr_test}.

\section{Results}\label{sec:5}
In this section, we delve into the detailed analysis of the ability of \ac{EV} fleets to provide two distinct flexibility products: capacity limitations and redispatch. 

Firstly, we employed Monte Carlo simulations to estimate the probabilities of the ability to deliver flexibility products for different CS categories under varying aggregation levels. This analysis was performed for two flexibility request windows: 8:00-11:00 (morning peak) and 18:00-21:00 (evening peak), which were chosen for generally high loading (and therefore potential congestion) in distribution networks. 
Randomly selected charging stations and days, according to Algorithm \ref{mc_dispatch_alg}, were utilised to simulate a variety of charging scenarios.\par

Subsequently, we trained machine learning models to directly predict the redispatch flexibility products. The performance of these predictive models was then compared with the results obtained from the Monte Carlo analysis.\par

To initiate our analysis, we applied the dispatch strategies (\emph{unoptimised}, \emph{optimal}, and \emph{greedy congestion avoidance}) as explained in Section \ref{dispatch_stratgies_section}. We employed Monte Carlo analysis to assess the impact of these strategies on different categories of \ac{CS} and various levels of aggregation.\par

For each category of \ac{CS} (shared, commercial, and residential), a total of 2000 samples were generated per discrete quantity of \ac{CS},  and for both the morning and evening peak hours. Incorporating a robust statistical approach, the Monte Carlo analysis results encompass error bars that effectively capture the sampling error. This was achieved by employing bootstrapping with 10,000 repetitions and plotting the bootstrap standard deviation.

\subsection{Redispatch flexibility product}
\begin{figure*}
     \centering
     
         \includegraphics[width=0.85\textwidth]{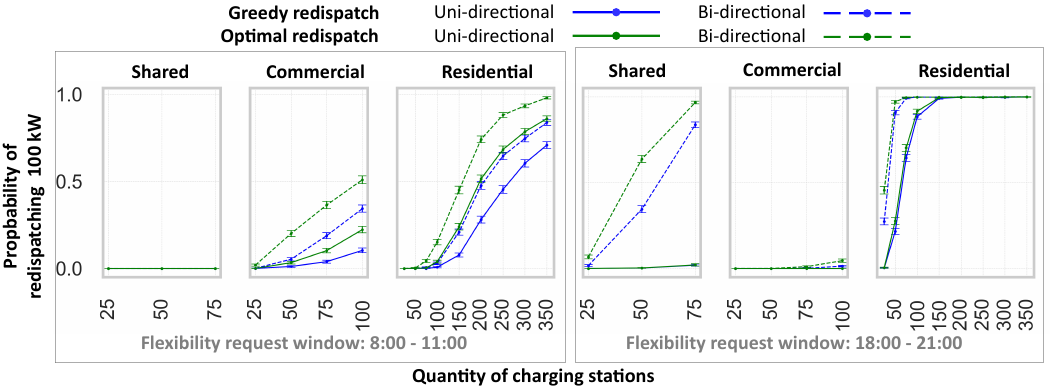}
        
        \caption{Probabilities of offering redispatch flexibility product for different variations in dispatch strategies, quantities of \acp{CS} and categories of \acp{CS}.}
        \label{fig:re_dispatch_res}
\end{figure*}
\figref{fig:re_dispatch_res} indicates the implied probability of the ability to deliver a redispatch product (at least 100 kW) based on the deterministic model for various aggregation levels and two flexibility request windows. During evening hours, residential chargers demonstrate a higher occupancy rate, enabling them to reliably provide at least 100 kW redispatch by aggregating 150 or more charging stations. In contrast, shared and commercial chargers cannot offer redispatch at this level. \par

In the morning flexibility request period, reduced \ac{EV} availability at residential locations significantly decreases the ability to offer the redispatch product. However, commercial locations perform better during this period due to the higher availability of \acp{EV} at workplaces. Shared charging locations exhibit similar trends, strongly correlated with the \ac{EV} connection probability (cf. \figref{fig:load_quantile}).  \par

Unoptimised charging (ref. Section \ref{dispatch_stratgies_section}) was used as the base scenario for all the above scenarios. The optimal dispatch strategy outperforms other approaches; however, its practical implementation is challenging due to the uncertainty in arrival times. To address this, a greedy congestion avoidance dispatch algorithm is simulated and compared with the optimal redispatch scenario. The greedy redispatch estimates a conservative shift from \ac{BAU} by calculating the minimum difference between the unoptimised aggregate power profile and the greedy dispatch aggregate power profile within the flexibility request window. Finally, as expected, switching from unidirectional to bi-directional charging significantly increases performance for both charging strategies.\par

Although the results focus on a threshold of 100~kW (the minimum redispatch bid size under current Dutch regulations), similar trends are observed for other thresholds: increasing the number of \ac{CS} in each category enhances the probability of achieving a given threshold power. Conversely, for a fixed number of \ac{CS}, the likelihood of reaching a redispatch threshold decreases as its power increases.

\subsection{Capacity limitation flexibility product}
\begin{figure*}
     \centering

         \includegraphics[width=\textwidth]{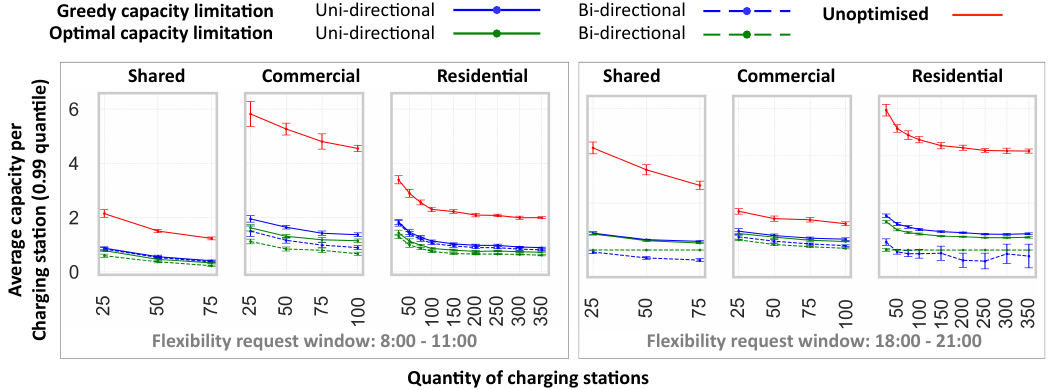}
     
        \caption{The maximum capacity per \ac{CS} for various flexibility request windows, illustrated for 99\% quantiles while considering random quantities of \ac{CS} across different categories (shared, residential, and commercial).}
        \label{fig:capacity_limitation}
\end{figure*}
On-street \ac{EV} chargers typically have individual connection agreements for the maximum supported charging power, up to 22 kW for newer three-phase AC chargers. As a result, the aggregate connection capacity for a single \ac{CPO} with distributed \acp{CS} is very large. In practice, only some \acp{CS} are active at any given time, making the aggregate connected capacity much lower than the available connection capacity -- and this can be further reduced with smart charging. To leverage this flexibility of connection capacities, the capacity-limitation product, as outlined by the Netherlands' congestion management regulations, can be lucrative for \acp{CPO} and beneficial for network operators. 

\figref{fig:capacity_limitation} shows the required capacity per \ac{CS} (i.e. the maximum recorded average power) for different dispatch strategies. Each sampled set of charging sessions results in a different capacity requirement, resulting in a distribution of capacities. The results plotted here are for 99\% quantiles. However, similar trends are observed for other quantiles. The diversified load requirement for fifty \acp{CS} does not exceed 6~kW for unoptimised charging. Moreover, compared to unoptimised charging, the reduction in the case of smart charging is considerable for both morning and evening flexibility request windows. 

The most significant and consistent drop can be seen for the residential charging locations due to the availability of a large fleet and high connection probability during evening hours. As the commercial charging locations are seldom used during the evening hours,  their average capacity per \ac{CS} is already low for unoptimised charging compared with other categories. However, this category shows significant reduction potential during the morning hours.

Bidirectional charging can restrict the capacity per \ac{CS} below 1 kW for all cases with more than 50 \acp{CS}. As the shift from unoptimised charging to unidirectional smart charging is more than that of bidirectional charging due to unidirectional capacity limitation, bidirectional charging may not be very lucrative for this particular product, provided its cost and effect on \ac{EV}'s battery health is more than that of unidirectional charging. 

\subsection{Duration of flexibility request window}

\begin{figure*}
    \centering
    \includegraphics[width=0.85\linewidth]{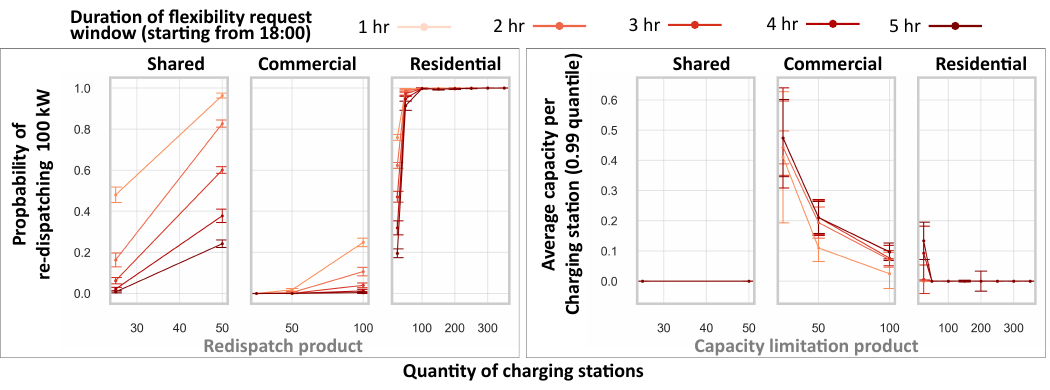}
    \caption{Effects of varying the duration of flexibility request window on the delivery of different flexibility products. The flexibility window starts from 18:00 hours during the evening peak. Results are plotted for different categories of \acp{CS} using an optimal dispatch scheme with bidirectional charging.}
    \label{fig:window_variation}
    
\end{figure*}

The span of the flexibility request window affects the magnitude of flexibility that could be leveraged for a given system. To demonstrate this, windows with different durations are simulated for redispatch and capacity limitation congestion products. As illustrated in \figref{fig:window_variation}, the ability to offer a specific product decreases with the increase in the length of the request window. This is due to the increase in the number of hours the combined \ac{EV} population has to shift with the increase in window length. The increase in window length will also cause a stronger rebound effect on the aggregate charging profile once the request window ends. After the flexibility request window, the rebound peak may be managed by adding a further incentive to spread charging across the available time until departure.

\subsection{Forecasting of flexibility products}

\begin{figure*}
     \centering
     
         \includegraphics[width=0.9\textwidth]{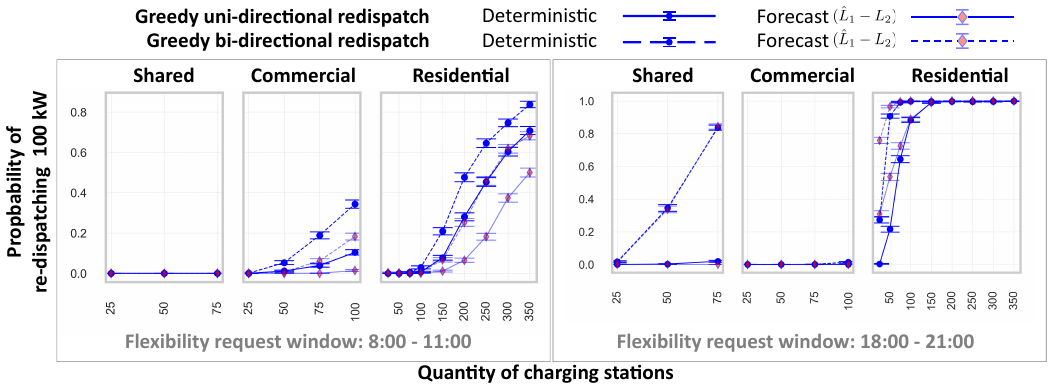}
 
        \caption{Comparison of redispatch using greedy avoidance strategy from forecasted day-ahead profile and the deterministic day-ahead profiles. Solid lines represent unidirectional charging, whereas the dashed lines show bi-directional charging. Forecasted and deterministic values are represented by diamond and circular markers, respectively.}
        \label{fig:model_compare_re_dispatch}
\end{figure*}

\begin{figure*}
     \centering
     \includegraphics[width = 0.9\linewidth]{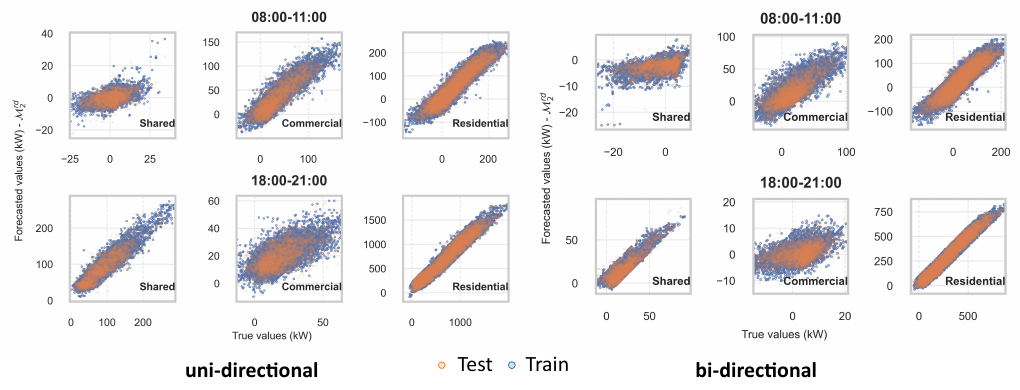}
        \caption{Model performances for forecasting redispatch operationally using $\mathcal{M}_2^{rd}$. Both training and test datasets are plotted.}
        \label{fig:model_performance_}
\end{figure*}

\begin{table*}[]
\caption{Forecasting model performances trained for different \ac{CS} categories, charging strategies and flexibility request window}
\label{table:performance}
\begin{tabular}{cc|cccccc|cccccc}
\hline
\multicolumn{2}{c|}{\textbf{CS categories}}                                                                               & \multicolumn{2}{c}{\textbf{Shared}}      & \multicolumn{2}{c}{\textbf{Commercial}}  & \multicolumn{2}{c|}{\textbf{Residential}}              & \multicolumn{2}{l}{\textbf{Shared}}                            & \multicolumn{2}{l}{\textbf{Commercial}} & \multicolumn{2}{l}{\textbf{Residential}}                       \\ \hline
\multicolumn{2}{c|}{\textbf{Charging strategy}}                                                                           & \textbf{Uni} & \textbf{Bi}               & \textbf{Uni} & \textbf{Bi}               & \textbf{Uni}              & \textbf{Bi}                & \textbf{Uni}              & \textbf{Bi}               & \textbf{Uni}   & \textbf{Bi}   & \textbf{Uni}              & \textbf{Bi}               \\ \hline
\multicolumn{2}{c|}{\textit{\textbf{\begin{tabular}[c]{@{}c@{}}Flexibility\\ request window\end{tabular}}}}               & \multicolumn{6}{c|}{\textit{8:00-11:00}}                                                                                                              & \multicolumn{6}{c}{\textit{18:00-21:00}}                                                                                                       \\ \hline
\multirow{2}{*}{\textbf{$\mathcal{M}_1^{da}$}} & \textit{\textbf{R2}}                                                     & 0.207        & 0.186                     & 0.59         & 0.762                     & 0.804                     & 0.865                      & 0.752                     & 0.832                     & 0.135          & 0.447         & 0.952                     & 0.942                     \\ \cline{2-14} 
                                               & \textit{\textbf{\begin{tabular}[c]{@{}c@{}}RMSE\\ (kW/CS)\end{tabular}}} & 0.117& 0.083& 0.082& 0.065& 0.044& 0.047& 0.058& 0.059& 0.085& 0.090& 0.042& 0.045                     \\ \hline
\multirow{2}{*}{\textbf{$\mathcal{M}_2^{rd}$}} & \textit{\textbf{R2}}                                                     & 0.176        & 0.217                     & 0.613        & 0.63                      & 0.879                     & 0.879                      & 0.618                     & 0.599                     & 0.32           & 0.323         & 0.895                     & 0.895                     \\ \cline{2-14} 
                                               & \textit{\textbf{\begin{tabular}[c]{@{}c@{}}RMSE\\ (kW/CS)\end{tabular}}} & 0.120& \multicolumn{1}{l}{0.132} & 0.082& \multicolumn{1}{l}{0.099} & \multicolumn{1}{l}{0.062} & \multicolumn{1}{l|}{0.061} & \multicolumn{1}{l}{0.089} & \multicolumn{1}{l}{0.091} & 0.127          & 0.123         & \multicolumn{1}{l}{0.058} & \multicolumn{1}{l}{0.058} \\ \hline
\end{tabular}
\end{table*}
The previous analysis assumed perfect information about \ac{EV} charging requirements at times of congestion. In reality, the \ac{CPO} must announce the ability to offer redispatch limitation products ahead of time based on a forecasted ability to deliver. It is, therefore, subject to additional uncertainty. To include uncertainty while estimating the probabilities of meeting the requirements for predicted flexibility products, their performance is compared with the deterministic case, using the workflow described in Section \ref{ML_work_flow_sections}\par

As explained, two models were trained and tested to predict \ac{BAU} and greedy redispatch for the three charging station categories, using two fixed flexibility request windows - one in the morning and one in the evening. The models were built using \acp{SVR} with a Gaussian kernel. Table:\ref{table:performance} and \figref{fig:model_performance_} illustrate the performance of the trained model for both test and train data.\par

\figref{fig:model_compare_re_dispatch} demonstrates the ability to redispatch at least 100 kW in two cases: (i) redispatch from the deterministic day-ahead profile, and (ii) redispatch from the forecasted day-ahead profile by model $\mathcal{M}_1^{da}$.  The ability to reliably provide redispatch ability over 100~kW is reduced by relying on day-ahead forecasts. This gives us a conservative estimate of the potential flexibility the \ac{CPO} can bid. Nevertheless, dependable delivery of this product is still possible for the evening peak, either using residential \acp{CS} or shared \acp{EV}, and bidirectional charging abilities further enhance this. \par

 To enable operational forecasting of redispatchable power without requiring such detailed data, the model $\mathcal{M}_2^{rd}$ can be used, provided it is properly tuned to minimise false positives. Avoiding false positives is essential, as forecasting that a \ac{CPO} can reliably offer a certain redispatch threshold when it actually cannot may lead to substantial penalties and a consequent reduction in overall revenue. \figref{fig:model_performance_} presents the performance of the trained model $\mathcal{M}_2^{rd}$ on both training and test datasets. The model can predict the redispatch ability qualitatively. Depending on the \ac{CPO}’s risk preference, the training procedure could be modified to penalise either over- or under-prediction.

\section{Conclusion}\label{sec:6}
This study addresses the congestion-related challenges faced in the power system due to aggregate uncontrolled charging of \acp{EV} and the potential to use smart charging to reduce congestion.
The ability to dependably offer two types of congestion management products (capacity limitation and redispatch) using various aggregations of \acp{EV} is quantified using a model based on historical charging transactions and product specifications in line with the Netherlands' congestion management regulations. In addition to (partially) clairvoyant optimisation approaches, machine learning models were used to directly predict the redispatch flexibility offering, including uncertainties associated with individual charging sessions.\par

It is seen that the ability to offer flexible products varies based on the level of aggregation, dispatch strategies and location-specific usage patterns of the \acp{CS}. Three main categories of \ac{CS} were identified and used for the analysis based on the characteristics of charging sessions. Among the discussed strategies, optimal (clairvoyant) dispatch performs the best, followed by greedy dispatch, which requires no information on future arrival times of \acp{EV}. However, when looking at aggregate performance, they are remarkably similar. We also examined all models for the case of bi- and mono-directional charging. As a general trend, bi-directional charging outperforms mono-directional charging in performance.\par

Looking at capacity limitations, the first finding is that there is already substantial `natural' diversity with unoptimised charging, so the peak power for aggregate \acp{CS} is significantly reduced from the 22~kW peak power of a single \ac{CS}. Significant further reductions can be achieved using smart charging. As a result, network operators should consider the baseline diversity and the additional gains from smart charging in their capacity planning. \par

Regarding redispatch products, the main practical barrier is the minimum redispatch quantity of 100 kW. In our study, only the residential \acp{CS} could dependably reach this threshold for the evening window. This is due to their number (in the data set) and the higher availability of these vehicles during evening hours. Machine learning models performed well in forecasting market offerings for congestion management services, highlighting their value for aggregators and network operators.

In conclusion, the study presented a comprehensive analysis of the aggregate flexibility potential of \acp{CS} for offering dependable congestion management products, focusing on a Dutch case study. By analysing over 500 thousand real \ac{EV} charging sessions, the research provided valuable insights into the challenges and opportunities posed by the growth of \acp{EV} and the implementation of smart charging flexibility within the Netherlands' energy landscape. Despite being grounded in the Dutch regulatory and operational context, the study offers a generalisable framework for assessing flexibility potential, which can serve as a reference for other regions facing similar grid constraints. The findings support the role of EV charging infrastructure as a practical and scalable source of flexibility in distribution grids, contributing to the broader goals of energy transition and grid resilience. \par

Considerations for future research include the effectiveness of redispatch products relative to different base profiles, such as cost-minimised or emission-minimised charging. Moreover, network congestion time can vary based on other factors such as weather, seasons or socioeconomic conditions. Hence, it is also essential to analyse the implications of different flexibility request windows on the potential of delivering congested products.\balance

\section{Acknowledgement}
The authors acknowledge the use of computational resources of the DelftBlue supercomputer, provided by Delft High-Performance Computing Centre~\cite{DHPC2022}. The authors would also like to thank the ROBUST consortium partners for fruitful discussions during the preparation of this paper.

\bibliographystyle{ieeetr}

\bibliography{references_final.bib}

\end{document}